# Coherent Ferroelectric Switching by Atomic Force Microscopy

A. Yu. Emelyanov*

*A. F. Ioffe Physico-Technical Institute, Russian Academy of Sciences, 194021 St. Petersburg, Russia*

General energy approaches have been applied to study the single-domain polarization reversal induced by the voltage-modulated Atomic Force Microscopy (AFM) in ferroelectric single crystals and thin films. Topographic analysis of energy surfaces in the subspace of domain dimensions is performed, and energy evolutions under an external bias are elucidated. This has let to successfully describe all stages of the AFM switching, including formations of a reversed domain, its growth in a bulk, domain contact instabilities near an electrode, and pure sidewise expansions in a film.

Recent progress in applications of Atomic Force Microscopy (AFM) to explore ferroelectrics at nanoscales has revealed new opportunities to address fundamental problems of switching phenomena, domain dynamics, and size effects [1]. Visualization of fine domain structures, local hysteresis measurements, and the writing of stable dots with 30 Gbit/ $cm^2$ density are among the remarkable AFM achievements on ferroelectric materials [1-3]. However, our understanding of repolarization processes induced by a voltage-modulated AFM is not free from serious drawbacks. No consistent model of polarization reversals has been proposed for the usual AFM setup when the switching is carried out within a low-voltage regime 1-50V. The general approach is to analyze domain energetics, but it has been applied only to interpret AFM data on bulk crystals at kV loads [4]. A theory for the AFM switching in thin films has not been introduced at all, though reliable predictions for the films are required to exploit the AFM reading/writing in memory devices. Thus, an appropriate model should be developed for explaining the AFM switching phenomena (1) at low voltages and (2) for ferroelectric thin films.

To get agreement between theories and AFM data is also important for resolving the main controversy in understandings of a macroscopic switching. A classical energy approach has predicted [5] that the nucleation of a reversed domain in uniformly polarized materials is impossible at the electric fields, which are routinely used for the actual switching of ferroelectric capacitors. This paradox implies that some features of real ferroelectrics like defects, surface states, or residual non-switched domains should be regarded as possible seeds in repolarization processes. A few efforts [6] have been undertaken to overcome the controversy, but their validity is not proven experimentally and remains questionable. On the other hand, the AFM opens the way to study a single-domain repolarization in small defect-free volumes and may be helpful to identify the factors, which significantly influence the macroscopic switching. The domain nucleation in AFM is expected to occur just under the tip apex, and domain growth is controlled by highly inhomogeneous electric fields concentrated near a tip-sample contact area. Such predefined dynamics may be referred to as a *coherent switching* contrary to the statistical macroscopic switching, where the random nucleation and/or multi-domain repolarization take place.

In this Letter, a model of the ferroelectric switching by the voltage-modulated AFM is proposed both for bulk crystals and for thin films. It is shown the AFM-induced polarization reversal is described by the classical energy approach in agreement with experiments at low voltages. The theory is based on calculations of electric fields via image-charge methods and properly involves the depolarization-field effects for thin-film geometry.

Let us imagine that a ferroelectric domain of semiellipsoidal shape is extending into an oppositely polarized *half-space*, both having the spontaneous polarization $\mathbf{P_s}$ orthogonal to a plane free surface. Let the out-of-plane component $E_z$ of an applied electric field $\mathbf{E}$ is parallel to $\mathbf{P_s}$ inside the domain. The energy change $\Delta W$ due to domain formation is determined by competitions between the electric-field energy, which favors to enlarge the domain, and the sum of domain-wall and depolarization contributions hindering the domain growth [5]. Assuming the in-plane dielectric isotropy as in $PbTiO_3$ crystal with the out-of-plane orientation of its polar *c* axis, we may consider the domain as half of the spheroid having a radius $r$ on the free surface and another semi-axis (length) $l$ along the surface normal. Counting radial $\rho$ and axial $z$ cylindrical coordinates from the spheroid center into a ferroelectric crystal, the energy $\Delta W$ can be written as

$$\Delta W = F\frac{16\pi^2 P_s^2}{3\varepsilon_c} r^2 l + C\frac{\gamma}{2} - 4\pi P_s \int_0^r \rho d\rho \int_0^l |E_z| dz. \quad (1)$$

The depolarization / field energy is defined by the first / last term in Eq. (1), the surface contribution is given via the spheroid surface area $C$ and specific energy $\gamma$, and $F$ denotes the depolarization factor of the spheroid along $z$ axis. $C$ and $F$ are analytical functions of $r$ and $l$ [7,8], the factor $F$ in anisotropic media being expressed via $l_a \equiv l\sqrt{\varepsilon_a/\varepsilon_c}$ instead of $l$ with the relative in-plane $\varepsilon_a$ and out-of-plane $\varepsilon_c$ permittivities of the crystal [5]. If the half-space is now converted to a *film* of thickness $H$, another depolarization term $W_f$ should be added to the right-hand side of Eq. (1). Depolarization-field calculations are reminiscent of image methods to find the field from a charge in a film [9], but now a charged domain boundary is reflected from film surfaces as a whole. If the domain grows from the top film surface, the depolarization-energy problem is reduced to summing electrostatic interactions of the charged half-spheroid with the images having their centers at distances $L_k = 2kH\sqrt{\varepsilon_a / \varepsilon_c}$ ($k$ =1,2,...) up and down from this surface. Then, the term $W_f$ is given by the set ($i$ =1,2,3):

$$W_f = \frac{16\pi P_s^2}{\sqrt{\varepsilon_a \varepsilon_c}} \frac{r^3}{e^3} \sum_{k=1}^{\infty} [-F(A_{1k}) + 2F(A_{2k}) - F(A_{3k})],$$

$$A_{ik} = (x+y)^2 + \frac{l_a^2}{r^2}\left(\frac{eL_k}{l_a} \pm \sqrt{e^2 - x^2} \pm \sqrt{e^2 - y^2}\right)^2,$$

and $$F(A_{ik}) \equiv \int_0^e x dx \int_0^e \frac{y dy}{\sqrt{A_{ik}}} K\left(\sqrt{\frac{4xy}{A_{ik}}}\right). \qquad (2)$$

Here $K(x)$ is complete normal elliptic integral of the first kind, $e = \sqrt{1 - r^2 / l_a^2}$ at $r < l_a$, and $e = \sqrt{1 - l_a^2 / r^2}$ at $r > l_a$. The signs in $A_{ik}$ alternate as follows: two minuses ($i$ =1); plus and minus ($i$ =2); both pluses ($i$ =3). Representing $K(x) = \pi \sum_{j=0}^{\infty} (a_j x^{2j}) / 2$ with $a_0$ =1 and $a_j = a_{j-1}(1 - 0.5/j)^2$ [7], the domain energetics in the film is described by Eqs (1)-(2) self-consistently [10].

To define the electric field $\mathbf{E}$ under a biased tip, multi-charge procedures have been developed for the AFM on a metal and dielectric crystal / film [11]. They give an equal electric potential on a surface of conducting tips by exploiting the known image-charge algorithms [8] but with *many* initial charges inside the tip. Previous attempts to model electric fields in AFM by a single charge [2-4] cannot reproduce the equipotential tip surface. To satisfy with this condition is extremely important for the AFM contact modes, which are used in switching experiments and characterized by the relation $D << R$ between a tip-sample distance $D$ and a tip radius $R$ of curvature. The single-charge approaches are also inadequate to map electric fields in thin films with $H \sim R$.

Consequently, no correct description of the switching can be achieved by these models at low voltages contrary to the multi-charge method used below. In field calculations, the tip is treated as the circular cone of height $h$ and half-angle $\theta$ terminated by a sphere of radius $R$, and the tip surrounding is characterized by a dielectric constant $\varepsilon_s$. Two sets $h$=10 μm, $\theta=15^0$, $R$=10 nm (tip $A$) and $h$=10 μm, $\theta=35^0$, $R$=50 nm (tip $B$) are specified, and a point contact ($D=0$) of the tip with a free ferroelectric surface in vacuum ($\varepsilon_s$=1) is assumed. We mainly consider lead titanate at $25^0$C with $\varepsilon_a$= 123, $\varepsilon_c$=66, $P_s$=0.757 C/m$^2$, and $\gamma$=0.169 J/m$^2$ [12].

After integrating the electric field in Eq. (1), we are able to quantify the energy $\Delta W$ as a function of the domain sizes $r$ and $l$. A typical energy surface on the $(r,l)$-plane is exemplified on Fig. 1 for $PbTiO_3$ bulk crystal at the applied voltage $U$=2 V. The main features of the functional $\Delta W(r,l)$ are its saddle point $S$ with an altitude $W_c$ at a point $(r_c,l_c)$, as well as the (global or local) minimum $M$ with a value $W^*$ and nonzero coordinates $(r^*,l^*)$. The magnitude $W_c$ gives the barrier (*activation energy*) for domain nucleation, and the *critical sizes* $r_c$ and $l_c$ in $S$ distinguish two parts in all polarization states. Domains of dimensions $r<r_c$ and $l<l_c$ are unstable and disappear within atomic-relaxation times [Fig. 1(a)], whereas those with $r>r_c$ and $l>l_c$ turn into a stable or metastable state in $M$ [Fig. 1(b)]. The stable domain (in applied field!) has the energy $W^*$<0 (global $M$), and $W^*$>0 for the metastable state (local $M$). The maximal repolarized volume, which can be attained in ideal nonconductive crystal at fixed $U\neq 0$, is defined by the *equilibrium sizes* $r^*(U)$ and $l^*(U)$.

The AFM switching is controlled by the applied voltage $U$, and the evolution of the points $S$ and $M$ with $U$ is of primary interest. Calculations yield the increase of $U$ leads to the decrease of activation parameters $W_c$, $r_c$, and $l_c$, but the equilibrium values $|W^*|$, $r^*$, and $l^*$ increase under enhanced voltages (Figs. 2-3). Thus, an external bias of suitable polarity promotes the polarization reversal both by accelerating thermal-fluctuation mechanisms and by increasing the volume of a reversed domain. From topographic analysis, the important voltages $U_{cr}$ and $U_{th}$ are established, which identify three different regimes in the switching process. For $U < U_{cr}$ with the *critical voltage* $U_{cr}$, $\Delta W$ is an increasing function of $r$ and $l$ so that repolarizations are impossible because of unsurmountable barriers $W_c$ [Fig. 2(a)]. A bias within $U_{cr} \leq U \leq U_{th}$ corresponds to a metastable state ($W^*$>0) when a reversed domain may be created but must be destroyed by fluctuations. If $U$ exceeds the switching *threshold* $U_{th}$, the repolarization in some volume is energetically preferable ($W^*$<0), and the reversed domain should exist until the bias $U$ is removed (Fig. 3). For $\mathrm{PbTiO_3}$ bulk crystal, $U_{cr}$=0.99 V, $U_{th}$=1.05 V (tip $A$) and $U_{cr}$=1.4 V, $U_{th}$=1.6 V (tip $B$). Figure 2(a) show the activation barrier is negligible for sharp tips ($A$) at $U > U_{th}$, whereas, for blunted tips ($B$), the nucleation may be a *rate-limited* stage in the overall switching process at low voltages. Indeed, a time $\tau$ to fluctuate a domain of the critical volume $V_c = 2\pi r_c^2 l_c/3$ is roughly estimated as $\tau \approx \tau_0 (V_c / V_0) \exp(W_c / kT)$, where $\tau_0$ and $V_0$ are characteristics of usual excitations like phonons ($kT$=26 meV at $25^0\mathrm{C}$). This gives rise to the dependence of AFM repolarizations on the frequency $f$ of applied voltages. If $f$ is higher than a threshold value $f_{th}$=$1/\tau$, the switching is prohibited. The functions $f_{th}(U)$ evaluated at typical $\tau_0$=1 ps and the unit-cell volume $V_0$ are presented on Fig. 2(b).

The energy surfaces and voltages $U_{cr}$, $U_{th}$ are strongly dependent on the tip radius $R$, distance $D$, and dielectric ratio $\varepsilon_s / \sqrt{\varepsilon_a \varepsilon_c}$, since the electric-field distribution is significantly influenced by these parameters. For instance, the decrease of $D$ is equivalent to some increase of $U$ due to similar effects on the field intensity. Generally, all parameters, which enhance the field *concentration* in the sample volume of a few unit cells near the tip apex, have to facilitate the nucleation by reducing $W_c$, $r_c$, $l_c$. In turn, the domain volume is increased by choosing the parameters so that the *average* field grows in the same volume. Following these receipts, it is easy to predict the increase of equilibrium ($r^*$, $l^*$, $|W^*|$) and activation ($r_c$, $l_c$, $W_c$) characteristics with increasing $R$. There are also some *optimal* values of the dielectric ratio to accelerate the nucleation or maximize domain volumes. These trends may be suitable to optimize the switching by modifying AFM setups or samples used.

Simulations show the initial stages of the domain nucleation and growth at $l << H \sim R$ are the same for thin films and bulk crystals, and quantitative differences come from trivial field enhancements in a film electroded on its bottom plane. However, two additive features distinguish the film from half-spaces in respect to domain energetics, both are concerned with the underlying electrode. First, at biases $U < U_j$ with a *jump* value $U_j$, the energy surface in the films contains a bottom saddle point $(r_b, l_b)$ with an altitude $W_b$ at $0.5H << l_b < H$ [13]. Second, a film minimum is brought onto this surface, and it is attained for a *cylindrical* domain penetrating the whole film thickness $H$. With increasing $U$, the bottom barrier $W_b$ is decreased and the film minimum is deepened. At $U_{cr} < U < U_j$, the energy surface includes three minims delimited by two saddle points, and the competition between the *bulk* (spheroids with $l < H$) and *film* (circular cylinders of length $l = H$) domains is observed for $U_{th} < U < U_j$. If $U > U_p$ with some bias $U_p > U_{th}$, the cylindrical domain becomes energetically favorable, and a thermally activated domain jump to the bottom electrode with the consequent spheroid-to-cylinder transition may be expected for $U_p < U < U_j$. Finally, at $U \geq U_j$, the bottom barrier is absent so that the domain, after nucleating on a top surface, runs through the film rapidly and spontaneously transforms to a cylinder. For $PbTiO_3$ and $BaTiO_3$, the voltages to overcome the barrier $W_b$ by fluctuations for a time $\leq 10$ s are found to be very close to the values $U_j$. Thus, the jump-to-bottom process, being thermally activated and frequency-dependent in nature, is mainly affected by the magnitude of applied biases, and only the jump voltages are of importance. For $PbTiO_3$ film of thickness 100 nm, $U_j \approx 32$ V (tip $A$) and $U_j \approx 12$ V (tip $B$).

The results obtained for a ferroelectric film mean the *contact instability* of AFM-written domains in two respects. First, there is no stable curved domain having its apex in the close vicinity of a continuous electrode. Second, curved domain boundaries touching an electrode tend to unbend. The reason of these effects is that the depolarization energy is decreasing (a) when the domain-spheroid is approaching the bottom electrode and (b) bound charges are compensated by free carriers in the domain-electrode contact area. The latter condition is fulfilled naturally, and the former is confirmed by Eqs. (1)-(2). A leading role of depolarization fields in domain dynamics is stressed by a considerable increase of the domain radius when the spheroid-to-cylinder transition is occurred. Indeed, the depolarization energy vanishes for the domain-cylinder in static conditions, and the total energy is simply the sum of field and wall ($\pi\gamma r\sqrt{r^2+H^2}$) contributions, where $r$ is now a cylinder radius. The *abrupt* increase of $r$ due to cancellation of depolarization fields is pronounced if the compensation of polarization charges on the top surface is provided, e. g., by retaining the bias $U$ near $U_j$ for sufficient time. This is seen from Fig. 3, where the equilibrium domain sizes for $PbTiO_3$ thin film are given as functions of $U$.

All previous analysis implies that (i) the polarization charges on crystal/film surfaces are compensated, but (ii) the bound charges on a domain boundary inside a ferroelectric bulk are not. Both conditions justify Eqs. (1)-(2), where the depolarization energy is given for a plate-capacitor setup and no bulk migration of free charges is permitted. For any stable domain-spheroid in $PbTiO_3$ film of thickness ~100 nm, we estimate the time $10^{-12}$-$10^{-8}$ s to satisfy with (i) by a charge transfer from the tip. In turn, (ii) is mainly violated by migrations of bulk charged defects. For oxygen vacancies as the most mobile species in titanates [14], the shortest interval ~$10^3$ s is obtained to remove discontinuity of $\mathbf{P_s}$ on the stable domains in the film interior. A time of AFM switching experiments is typically within $10^{-7}$-$10^{2}$ s, hence, the two conditions are fulfilled for a *forward* domain growth towards the bottom electrode. However, a *sidewise* expansion of the cylindrical domains after jumps-to-bottom is accompanied by the increase of surface depolarization fields, since strong reductions of the applied electric field outside the tip-sample contact lead to violation of (i). Evaluations yield the fields on the sample surface may be approximated as $\sim 1/\rho$ at $\rho >\sim 0.1R$. If charge mobilities are field-independent and a rate $dr/dt$ of the sidewise growth is controlled by the charge transport from the tip, a time $\Delta t$ to enlarge the domain-cylinder to a radius $r$ is $\Delta t \propto r^2$. Another mechanism to promote sidewise motions via generation of repolarized steps on a domain wall has been introduced by Miller and Weinrich [15]. It gives $\Delta t \propto \exp r$. Dependences close to $r \propto \sqrt{\Delta t}$ and $r \propto \ln \Delta t$ are observed in pulse AFM experiments [3] with $\Delta t$ as the pulse width. We may assume that the charge transfer from AFM tips determines the rate of sidewise expansions at $r <\sim R$, whereas Miller-Weinreich mechanism dominates for larger domains. Evidently, at $r > R$, the fields are too low to move a domain boundary as a whole, since a pinning by defects and even an atomic-lattice barrier become substantial [5]. In this case, the thermal activation of repolarized steps is the only driving force to further enlarge a domain.

Thus, the AFM switching involves *four stages* in *c*-oriented thin films of perovskite ferroelectrics ($PbTiO_3$, $BaTiO_3$, etc.): (1) Nucleation. It is frequency-independent for sharp tips in ideal contact conditions and occurs for biases $U > U_{cr}$. (2) Bulk growth. This is the forward domain motion with the minor sidewise expansion. The switching starts at $U > U_{th}$, and domain sizes are quasi-linear functions of $U$ not close to $U_j$. (3) Contact instability. It gives rise to the jump-to-bottom at $U \approx U_j$, degeneration of domain curvature, and a momentary increase of the repolarized volume.(4) Sidewise growth. It is the thermally activated domain expansion impeded by the surface depolarization fields and pinning barriers. The stages (1)-(2) are also typical for bulk crystals operated at low-voltages $U <\sim 100$ V. If the bias is turned off on these stages, an unpinned domain disappears.

The coherent switching is very sensitive to AFM setups, measurement techniques, contact conditions, and material properties so that all relevant parameters should be specified for accurate comparison with theory. Our predictions are in good accordance with various AFM data dispersed in the literature. To improve the model, piezoelectric distortions, dielectric relaxation, and pinning effects will be analyzed further. A good deal of work remains to be done in theory and experiment to reliably include AFM switching operations in nanoelectronics.

The author gratefully acknowledges N. A. Pertsev for stimulating discussions on AFM field problems, A. Ankudinov for helpful co-analysis of experimental results, and S. Emelyanova for assistance in simulations.

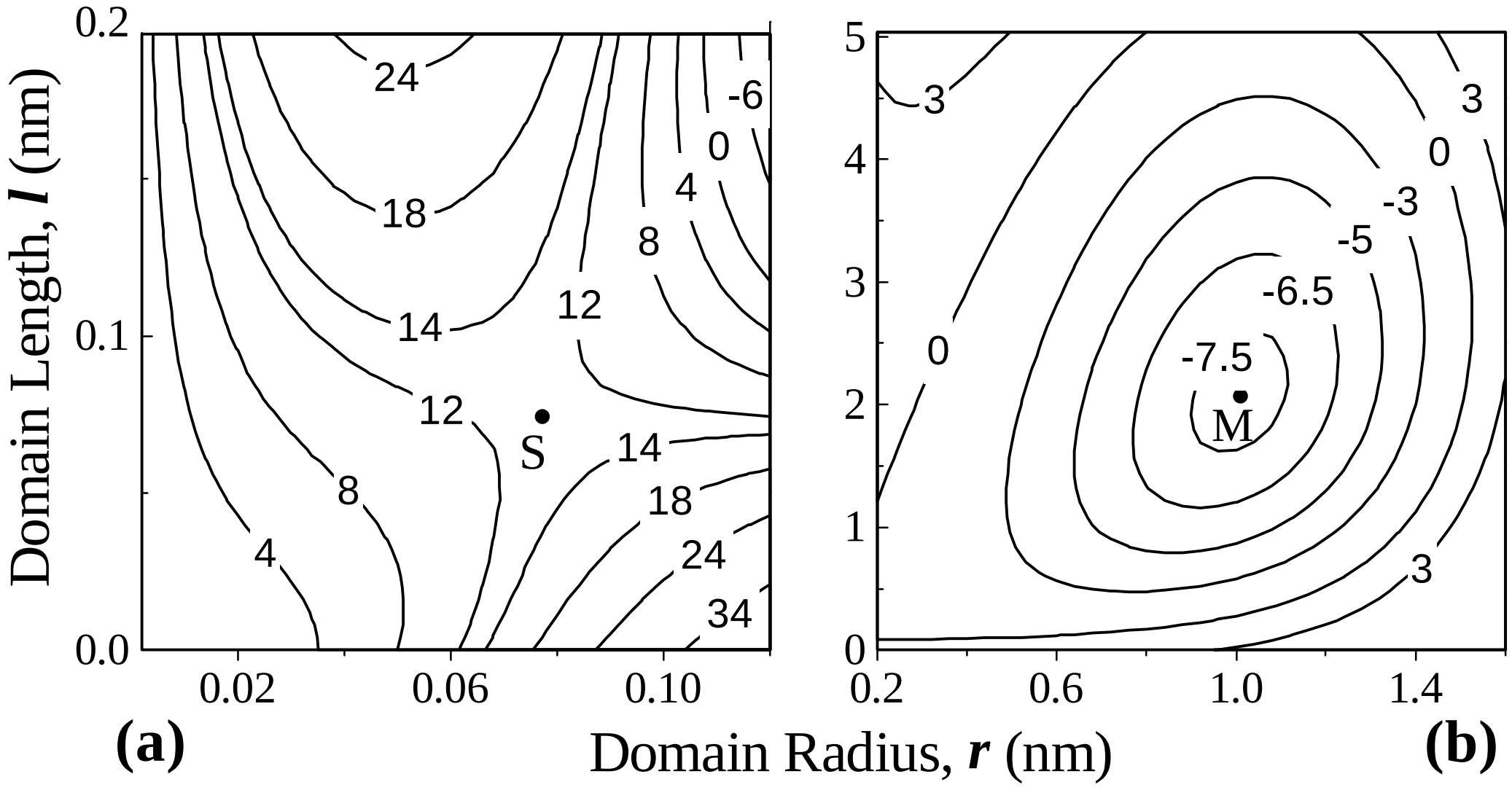

FIG. 1. Dependences of the energy $\Delta W$ on domain sizes $r$ and $l$ for $PbTiO_3$ single crystal (top views). Energy contours are plotted (a) near the saddle point $S$ and (b) close to the minimum $M$ at the bias $U$ =2 V on the tip $A$. Values in meVs (a) and eVs (b) are denoted on each curve.

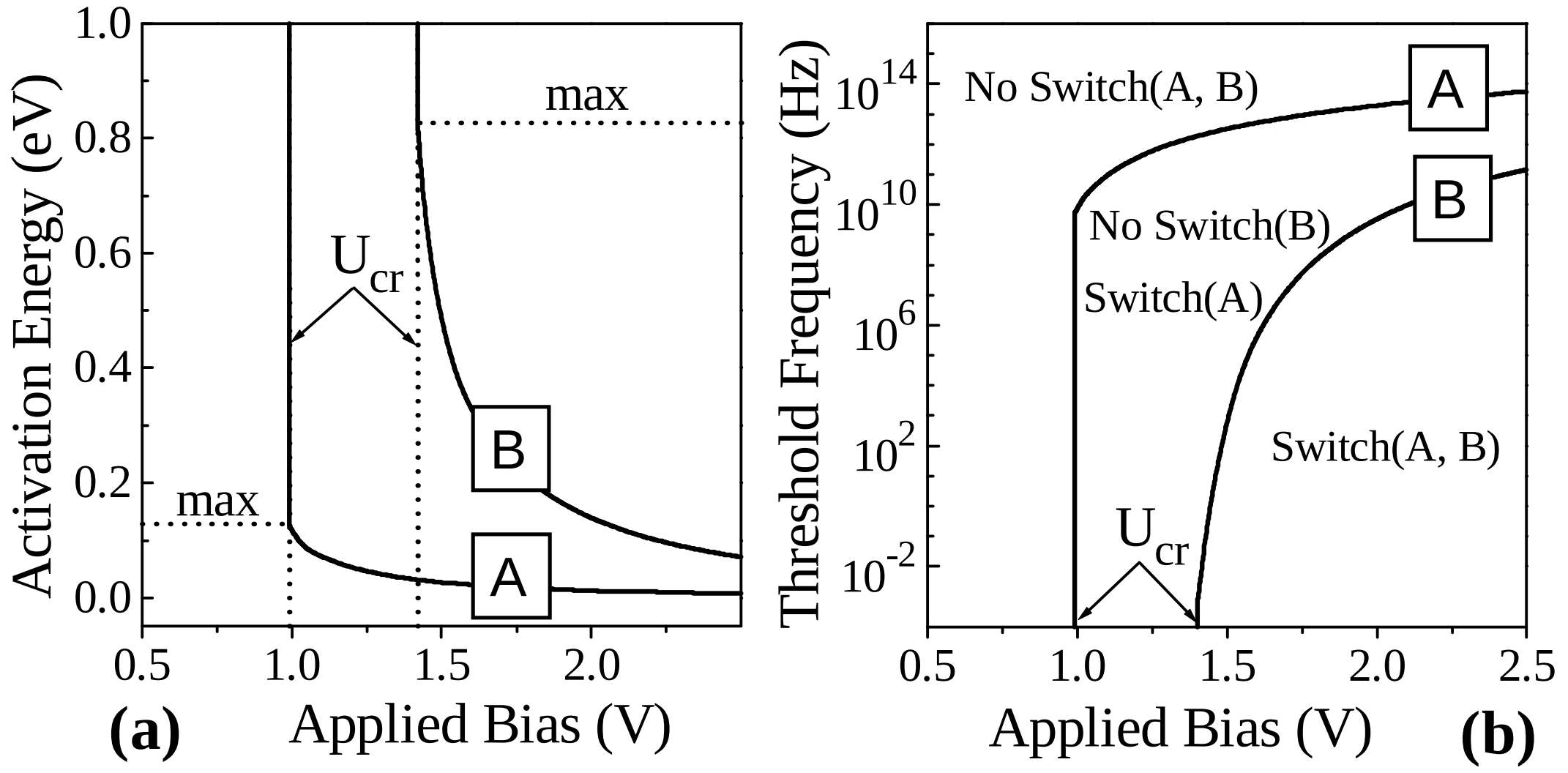

FIG. 2. The activation energy $W_c$ (a) and threshold frequency $f_{th}$ (b) as functions of the bias $U$ on tips $A$ and $B$ for $PbTiO_3$ bulk crystal. Critical voltages $U_{cr}$, maximal finite barriers in (a), and switchable regions in (b) are shown.

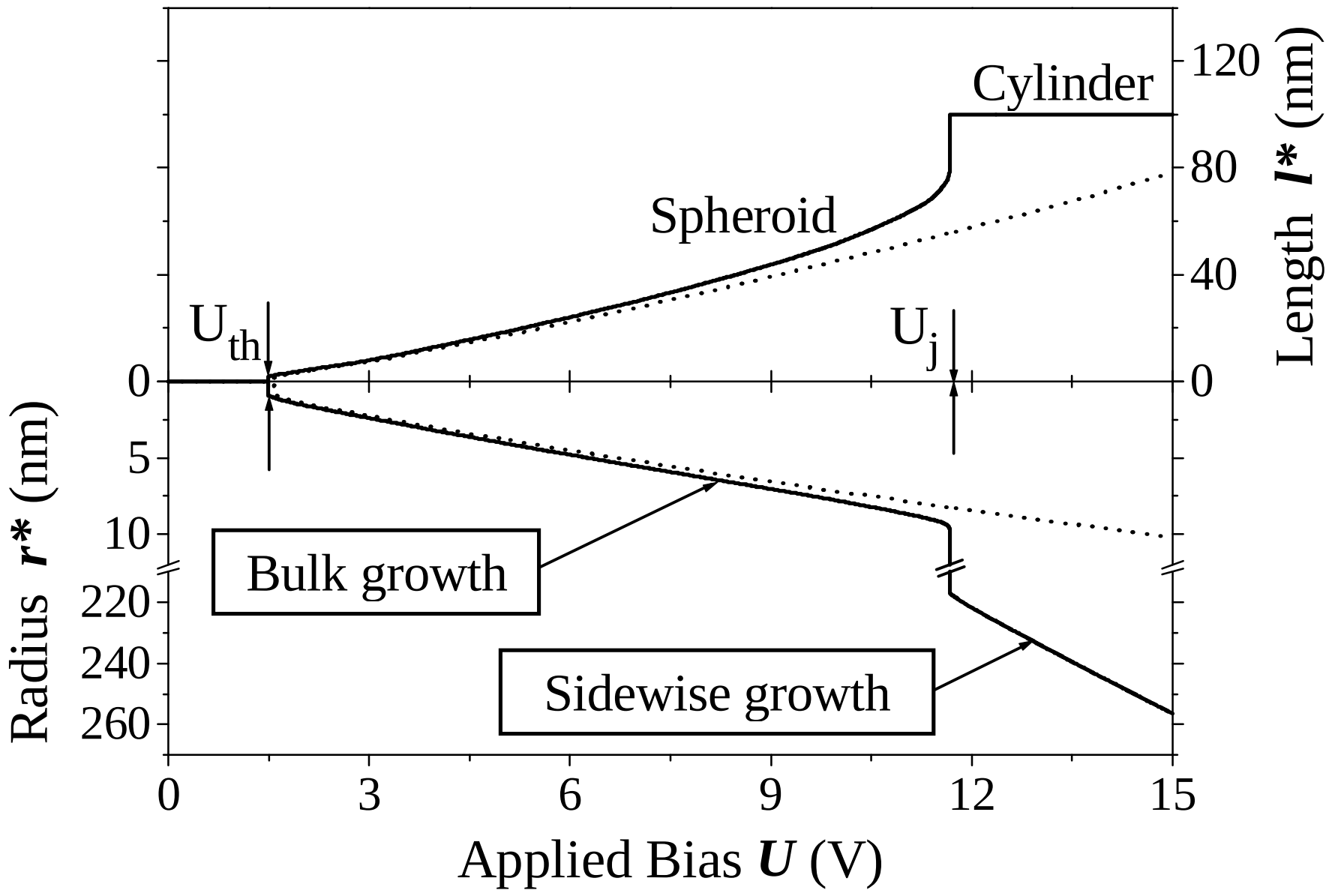

FIG. 3. Equilibrium domain sizes $r^*$ and $l^*$ versus the voltage on the tip $B$ in $PbTiO_3$ bulk crystal (dots) and thin film with $H$=100 nm (solid lines). For the film, the values $U_{th}$ and $U_j$, domain shapes, and. types of growth are given.